\documentclass[lettersize]{IEEEtran}
\usepackage{amsmath,amsfonts}
\IEEEoverridecommandlockouts
\normalsize

\usepackage{cite}
\usepackage{url}
\usepackage{hyperref}
\hypersetup{colorlinks=true, linkcolor=red, citecolor=blue, urlcolor=blue} 
\usepackage{amssymb}
\usepackage{amsmath}
\usepackage{array}
\usepackage{amsthm}
\allowdisplaybreaks 
\usepackage{autobreak}
\usepackage{mathptmx}
\usepackage{mathtools, cuted}
\usepackage{algpseudocode}
\usepackage[ruled, vlined, linesnumbered]{algorithm2e}
\usepackage{graphicx}
\usepackage{subcaption} 
\usepackage{bbm}
\usepackage{booktabs}
\usepackage{multirow}
\usepackage{makecell}
\usepackage[table,xcdraw,dvipsnames]{xcolor}
\usepackage{colortbl}
\usepackage{enumitem}
\usepackage[left=1.62cm,right=1.68cm,top=1.9cm]{geometry}

\allowdisplaybreaks

\newcommand{\rs}{\!\!}
\newcolumntype{C}[1]{>{\centering \arraybackslash}p{#1}}
\newcommand{\bblue}{\textcolor{black}}

\usepackage{acronym}
\acrodef{ml}[ML]{machine learning}
\acrodef{htransformer}[{\tt hTransformer}]{hybrid Transformer}
\acrodef{lstm}[{\tt LSTM}]{long short-term memory}
\acrodef{bs}[BS]{base station}
\acrodef{ue}[UE]{user equipment}

\acrodef{sgd}[SGD]{stochastic gradient descent}
\acrodef{cpu}[CPU]{central processing unit}
\acrodef{prb}[pRB]{physical resource block}
\acrodef{snr}[SNR]{signal-to-noise-ratio}
\acrodef{lp}[LP]{linear programming}
\acrodef{fpp}[FPP]{floating point precision}
\acrodef{cdf}[CDF]{cumulative distribution function}
\acrodef{uav}[UAV]{unmanned aerial vehicles}
\acrodef{ap}[AP]{access point}
\acrodef{hz}[Hz]{hertz}
\acrodef{csi}[CSI]{channel state information}
\acrodef{kkt}[KKT]{Karush–Kuhn–Tucker}
\acrodef{fifo}[{\tt FIFO}]{first-in-first-out}
\acrodef{trimtoplabel}[{\tt TrimTopLabel}]{trim top label}
\acrodef{fcn}[{\tt FCN}]{fully connected neural network}
\acrodef{lstm}[{\tt LSTM}]{long short-term memory}
\acrodef{bilstm}[{\tt BiLSTM}]{bi-directional long short-term memory}
\acrodef{cnn}[{\tt CNN}]{convolutional neural network}
\acrodef{fc}[{\tt FC}]{fully connected}
\acrodef{isac}[ISAC]{integrated sensing and communication}
\acrodef{pdp}[PDP]{power delay profile}
\acrodef{ber}[BER]{bit error rate}

\acrodef{agi}[AgI]{augmented information}
\acrodef{sfc}[SFC]{service function chain}
\acrodef{rt}[RT]{ray tracing}
\acrodef{mpc}[MPC]{multi-path component}
\acrodef{gscm}[GSCM]{geometry-based stochastic channel model}
\acrodef{ml}[ML]{machine learning}
\acrodef{ai}[AI]{artificial intelligence}
\acrodef{ddpm}[DDPM]{Denoising Diffusion Probabilistic Model}
\acrodef{genai}[GenAI]{generative AI}
\acrodef{pmp}[PMP]{pathloss map prediction}
\acrodef{gan}[GAN]{generative adversarial network}
\acrodef{rem}[REM]{radio environment map}
\acrodef{roi}[RoI]{region of interest}
\acrodef{dod}[DoD]{direction of departure}
\acrodef{doa}[DoA]{direction of arrival}
\acrodef{kpi}[KPI]{key performance indicator}
\acrodef{ssim}[SSIM]{structural similarity index measurement}
\acrodef{rms}[RMS]{root mean squred}
\acrodef{mse}[MSE]{mean square error}
\acrodef{rmse}[RMSE]{root mean square error}
\acrodef{mae}[MAE]{mean absolute error}
\acrodef{rms}[RMS]{root mean squre}
\acrodef{acf}[ACF]{auto correlation function}
\acrodef{sdb}[{\tt SDB}]{series decomposition block}
\acrodef{dd}[DD]{double-directional}
\acrodef{gpu}[GPU]{graphical processing units}
\acrodef{nlos}[NLOS]{non-line-of-sight}
\acrodef{los}[LOS]{line-of-sight}
\acrodef{vae}[VAE]{variational autoencoder}

\title{Double Directional Wireless Channel Generation: A Statistics-Informed Generative Approach} 
\author{
    \IEEEauthorblockN{Md Ferdous Pervej\IEEEauthorrefmark{1}, Patel Pratik\IEEEauthorrefmark{2}, Koushik Manjunatha\IEEEauthorrefmark{2}, Prasad Shamain\IEEEauthorrefmark{2}}, and Andreas F. Molisch\IEEEauthorrefmark{1}. \\
    \IEEEauthorblockA{\IEEEauthorrefmark{1}Ming Hsieh Department of ECE, University of Southern California, Los Angeles, CA, USA $90089$}\\
    \IEEEauthorblockA{\IEEEauthorrefmark{2}Amazon Lab$126$, Sunnyvale, CA, USA $94089$}\\
    \IEEEauthorblockA{Emails:  \{pervej, $\!\!$ molisch\}@usc.edu\IEEEauthorrefmark{1}, pratikp@lab126.com\IEEEauthorrefmark{2}, \{koushiam,prasadks\}@amazon.com\IEEEauthorrefmark{2} }
\vspace{-0.3in}
\thanks{The research of MFP and AFM was supported in part by a grant from Amazon.}
\thanks{\copyright $2025$ IEEE. Personal use of this material is permitted. Permission from IEEE must be obtained for all other uses, in any current or future media, including reprinting/republishing this material for advertising or promotional purposes, creating new collective works, for resale or redistribution to servers or lists, or reuse of any copyrighted component of this work in other works.}
}

\begin{document}

\maketitle
\IEEEpeerreviewmaketitle

\begin{abstract}
Channel models that represent 
various operating conditions a communication system might experience are important for design and standardization of any communication system.
While statistical channel models have long dominated this space, \ac{ml} is becoming a popular alternative approach. However, existing approaches have mostly focused on predictive solutions to match instantaneous channel realizations. 
Other solutions have focused on pathloss modeling, while \ac{dd} channel representation is needed for a complete description.
Motivated by this, we (a) develop a generative solution that uses a \ac{htransformer} model with a low-rank projected attention calculation mechanism and a \ac{bilstm} layer to generate complete \ac{dd} channel information and (b) design a domain-knowledge-informed training method to match the generated and true channel realizations' {\em statistics}.
Our extensive simulation results validate that the generated samples' statistics closely align with the true statistics while mostly outperforming the performance of existing predictive approaches.
\end{abstract}


\section{Introduction}
\acresetall 


The propagation channel is the heart of any wireless communication system and, thus, needs accurate modeling. 
We can, in general, distinguish between site-specific (one-to-one mapping of geometry to channel) models, which are mostly used for deployment planning, or completely statistical models (with many realizations) that represent a whole environment class and are often used for standardization purposes.
While these extreme cases are well explored and established it would be desirable to have an intermediate-type model that creates channels that can describe \emph{plausible} channel statistics and the evolution of these statistics in many cases. 
In particular, due to the dominant importance of multi-antenna systems, the statistics of the \ac{dd} channel representation \cite{steinbauer2001double}, which includes delay, power and angular information of different \acp{mpc}, are required.




Due to the great success of \ac{ml} in tackling various complex problems, it has been proposed for such tasks as channel prediction \cite{lee2024scalable} and channel modeling \cite{huang2022artificial,hu2024channel}.
The recent success of \emph{foundation models} such as Transformer \cite{vaswani2017attention}, which advocates for \emph{attention}-based learning, has remarkably increased the effectiveness of ML for many complex problems. 
The foundation models can be used for both predictive and generative tasks. 


Channel prediction, which are mostly used during operation for, e.g., beamforming and scheduling, is usually best for predicting instantaneous channel realizations.
However, as discussed above, for channel modeling, it is important to generate complete \ac{dd} channels that follow specific statistics for a long time.


In the generative paradigm, researchers have mainly used \ac{gan} for a long time as ``the" generative solution for wireless channel models \cite{yang2019generative,xiao2021channel,hu2024channel}. 
Xiao \textit{et al.} used location information to generate only the field strengths of different \acp{mpc} using \ac{gan} \cite{xiao2021channel}.
Hu \textit{et al.} also used \ac{gan} to generate directional channels in different mid-band frequencies \cite{hu2024channel}.
While \ac{gan} is popular for many generative applications, it is usually not suitable for sequential tasks where time dependency is crucial.  
\bblue{As a potential remedy, hybrid models with \ac{gan} and \ac{lstm} blocks are often used to capture the temporal dependencies \cite{li2022gan,huang2024frequency}.} 
Different from the above works, Huang \textit{et al.} used location information to generate delay, power and angular information using simple \ac{fcn} and radial basis function neural network \cite{huang2018big}.

Some recent works also leveraged Transformer-based architectures for channel prediction \cite{zhou2024transformer,jiang2022accurate,hua2024cross}.
Zhou \textit{et al.} predicted real and imaginary parts of \ac{csi} for multiple future time steps using vanilla Transformer \cite{zhou2024transformer}.
A similar prediction strategy and Transformer model was also used by Jiang \textit{et al.} \cite{jiang2022accurate}. 
Besides, Kang \textit{et al.} proposed a modified Transformer architecture and spatio-temporal attention calculation mechanism to predict multi-step \ac{csi} \cite{hua2024cross}.
These works used Transformer as a general predictor that tried to match instantaneous channel realizations and did not emphasize channel statistics.
Moreover, the prediction window is often very short (e.g., $5$ time steps \cite{zhou2024transformer,hua2024cross} and $5$ frames \cite{jiang2022accurate}), which may not provide any meaningful statistical information about the channel.


While the studies above have paved the way for data-driven channel modeling, no propagation knowledge-aided generative solutions for \ac{dd} channel generation emphasizing the statistics are currently available.
Our key contributions in this context are: (i) we, to the best of our knowledge, for the first time, propose how to generate a sequence of \ac{dd} channel realizations across realistic RX trajectories using a Transformer-based hybrid model that leverages the statistical properties of the channel for model training. 
(ii) we develop a new \ac{htransformer}, which is built upon the original Transformer \cite{vaswani2017attention} architecture with linear complexity attention calculation and short-term prediction benefits of \ac{bilstm} \cite{schuster1997bidirectional}.
(iii) we design a channel statistics-aided training method to generate complete \ac{dd} channel realizations to match ground truth statistics.
Our extensive simulation results suggest that the proposed statistics-aided training method can generate samples accurately matching the true statistics and is particularly beneficial for a large generation window.

\section{Preliminaries and Problem Statement}

Since the propagation channel largely depends on the location of the receiver (RX)\footnote{
In the following, we only consider the downlink and, thus, equate the RX with the (mobile) user equipment. 
However, the same channel representation applies to the uplink.} and the scatterers present in the environment, our goal is to model the \ac{dd} propagation channel for all possible trajectory points (of the same RX).

\subsection{Preliminaries} 
\subsubsection{Double-Directional Wireless Channel Model}


This work assumes a single transmitter (TX) is located at a fixed location ${\bf r}_{\mathrm{tx}}=\{x_{\mathrm{tx}},y_{\mathrm{tx}}, z_{\mathrm{tx}} \}$.
Denote the location of the RX by ${\bf r}_{\mathrm{rx}} = \{x_{\mathrm{rx}}, y_{\mathrm{rx}}, z_{\mathrm{rx}}\}$.
Given the ${\bf r}_{\mathrm{tx}}$ and ${\bf r}_{\mathrm{rx}}$, the \ac{dd} wireless channel has the following impulse response \cite{steinbauer2001double}
\begin{align}
\label{ddIR_Eqn}
    & h \left(t,\tau, \Omega,\Psi; {\bf r}_{\mathrm{tx}}, {\bf r}_{\mathrm{rx}} \right) \!= \!\!\sum\nolimits_{n=1}^{N(\mathbf{r})} \!\! |a_n| e^{j\phi_n} \delta (\tau - \tau_n) \delta(\Omega - \Omega_n)  \times \nonumber\\
    &\qquad \delta(\Psi - \Psi_n) e^{j2\pi \nu_n t} + h_{\mathrm{DMC}} (t,\tau, \Omega, \Psi; {\bf r}_{\mathrm{tx}}, {\bf r}_{\mathrm{rx}}),
\end{align}
where $t$, $\tau$, $\Omega$, $\Psi$, ${\bf r}_{\mathrm{tx}}$ and ${\bf r}_{\mathrm{rx}}$ are the time, delay, \ac{dod}, \ac{doa}, location of the TX and location of the RX, respectively. 
Besides, $N({\bf r})$ is the number of \ac{mpc} in that given location.
Furthermore, $a_n$, $\phi_n$, $\tau_n$, $\Omega_n$ and $\Psi_n$ are the gain, (random) phase, delay, \ac{dod} and \ac{doa} of the $n^{\mathrm{th}}$ path, respectively.
Moreover, $\nu_n$ is the Doppler shift\footnote{Note that for the case that TX and scatterers are static, the Doppler shift follows uniquely from the \ac{doa} and thus does not need to be separately modeled \cite{molisch2023wireless}.} and $h_{\mathrm{DMC}} (t,\tau, \Omega, \Psi; {\bf r}_{\mathrm{tx}}, {\bf r}_{\mathrm{rx}})$ is the diffuse \acp{mpc}.
It is worth noting that (\ref{ddIR_Eqn}) does not show dependency of $\Omega$, $\Psi$, $\tau$, $a$ on $t$ and ${\bf r}_{\mathrm{tx}}$ and ${\bf r}_{\mathrm{rx}}$ explicitly.

\subsubsection{Double-Directional Channel Statistics}




We discuss some widely used channel statistics below, which will also be used for our model training. 
The first statistic is \ac{rms} delay spread, which provides a measure of how the TX signal gets delayed in different \acp{mpc} before reaching the RX.
The \ac{rms} delay spread is calculated as
\begin{align}
    \rs\rs S_\tau = \sqrt{ \sum_{n=1}^{N(r)} \frac{|a_n|^2 }{\sum_{n'=0}^{N(r)} |a_{n'}|^2} (\tau_n - \bar{\tau})^2 } = \sqrt{\frac{\sum_{n=1}^{N(r)} |a_{n}|^2 \tau_n^2 }{\sum_{n'=0}^{N(r)} |a_{n'}|^2 } - \bar{\tau}^2},
\label{rmsDelaySpreadEqn}
\end{align}
where $\bar{\tau} \coloneqq \big[\sum_{n=1}^{N(r)} |a_n|^2 \tau_n \big] \big/ \big[ \sum_{n'=0}^{N(r)} |a_{n'}|^2 \big]$.

The second widely used statistic is the \ac{rms} angular spread, which provides information about angular variations in different \acp{mpc}.
While many calculate \ac{rms} angular spread following an analogous equation as in (\ref{rmsDelaySpreadEqn}), such definition may give rise to ambiguities due to $2 \pi$ periodicity of angles \cite{molisch2023wireless}.
As such, we use the definition of \cite{fleury2000first}
\begin{align}
    \rs\rs S_\Omega = \rs \sqrt{\Big[\sum\nolimits_{n=1}^{N(r)} \left\vert \exp{(j\Omega_n)} - \mu_\Omega \right\vert^2 \cdot \left\vert a_n\right\vert^2 \Big] \big/ \Big[\sum\nolimits_{n'=1}^{N(r)} |a_{n'}|^2 \Big]},
    \label{angularSpreadEqn}
\end{align}
where $\mu_\Omega = \big[\sum_{n=1}^{N(r)} \exp{(j \Omega_n)} \cdot \left \vert a_n \right\vert^2 \big]\big/\big[\sum_{n'=1}^{N(r)} \left\vert a_{n'} \right\vert^2\big]$.
Note that $S_\Omega \in [0,1]$ and is dimensionless \cite[Chapter $6$] {molisch2023wireless}.


In addition to the \acp{cdf} of those parameters, the \acp{acf} of the parameters are also important. 
While the current paper 
focuses on the basic principles and only considers the \acp{cdf}, 
the \acp{acf} will be considered in future work\footnote{\bblue{Modeling of \ac{acf} is not straightforward since it usually relies on the stationarity of the channel. 
However, the stationarity in different features of an \ac{mpc} does not necessarily hold over different periods. 
This becomes critical for \ac{ml} model training since the model usually gets trained over mini-batches, which only contain subsets of randomly sampled training data points from the entire training dataset.}}.

\subsection{Challenges and Limitations}

For the training and testing of our algorithms, one could consider \ac{rt} \cite{valenzuela1993ray}, which gives a complete representation of (\ref{ddIR_Eqn}). 
However, while it provides accurate delay, power and angular information of all \acp{mpc} for a given map and RX locations, results are typically available for each location as a list of \acp{mpc} that are ordered by power, making it difficult to track the evolution of each individual \ac{mpc} as the RX moves along a trajectory - in other words, associating which \ac{mpc} at a later location is the evolution of a particular \ac{mpc} earlier location is a difficult problem. 
Without this critical information, the \ac{ml} model may fail to capture the physics of \ac{mpc} evolution. 
An additional challenge is that the number of \acp{mpc} varies in \ac{rt} data. 
Since an \ac{ml} model requires an input/output shape, such variation is a major problem for \ac{ml} model training. 
While suitable preprocessing of \ac{rt} data might be able to overcome these problems, it might introduce ambiguities and errors. 
\bblue{Moreover, the use of real measurement data could be another option. 
Unfortunately, gathering a massive amount of such measurement data is time-consuming and very expensive.} 
Since our goal is the proof of principle of the statistics prediction and the performance assessment of the \ac{ml} algorithm (and not of the \ac{rt} preprocessing), we use a \ac{gscm} as a remedy. 
While such a model might not contain all the details of a real-world channel, it does represent the essential features \cite{molisch2004generic,molisch2023wireless}.

\subsection{Geometry-based Stochastic Channel Model}
\label{gscmSubSec}

In a basic GSCM, which we use here,  $N$ scatterers are placed. This placement is done according to a prescribed probability density function, but remains fixed during one simulation run (potentially with multiple trajectories).
Denote the $n^{\mathrm{th}}$ scatterer's location by ${\bf r}_{\mathrm{sc},n} = \left\{x_{\mathrm{sc},n}, y_{\mathrm{sc},n}, z_{\mathrm{sc},n}\right\}_{n=1}^{N}$. The power carried by each of the \acp{mpc} are computed by     
\begin{equation}
\label{pleqn} 
\begin{aligned}
    {\tt{PL} ~[in ~dB]} =& 13.54 + 39.08\log_{10}(d_{\mathrm{3d},{\bf r}_\mathrm{tx} \rightarrow {\bf r}_\mathrm{rx}}) + \\
    &\qquad\qquad 20 \log_{10}(f_c) - 0.6(h_{\mathrm{rx}} - 1.5)^2,
\end{aligned}
\end{equation}
where $d_{\mathrm{3d},{\bf r}_\mathrm{tx} \rightarrow {\bf r}_\mathrm{rx}}$ is the $3$-D distance (in meters) between TX and RX, $f_c$ is the carrier frequency (in GHz) and $h_{\mathrm{rx}}$ is the height of the RX. This model follows the urban macro path loss equation of the $3$GPP channel model \cite{3GPP38901}\footnote{While this model was derived from measurements to represent the {\em bulk pathloss}, not the pathloss for individual \acp{mpc}, we use it here for 
simplicity and because the details of the \ac{mpc} pathloss model have little impact on the algorithm behavior.}.
We assume the absence of a \ac{los} component: such a component might become the dominant contribution of the impulse response, and, thus, channel prediction and statistics would be reduced to describing the impact of this one component accurately.

For our further computation, the received power of the $n^{\mathrm{th}}$ path is written on a dB scale $g_n \coloneqq |a_n|^2 = 0 - {\tt{PL} ~ [in ~dB]}$, assuming TX power is $0$ dBm.
The $n^{\mathrm{th}}$ path's phase $\phi_n \in [-2\pi, 2\pi]$. Besides, we calculate the delay and angle information as  
\begin{align*}
    & {\tt Delay} \coloneqq \tau_n \coloneqq \big[ d_{\mathrm{3d}, {\bf r}_{\mathrm{tx}} \rightarrow {\bf r}_{\mathrm{sc},n}} + d_{\mathrm{3d}, {\bf r}_{\mathrm{sc},n} \rightarrow {\bf r}_{\mathrm{rx}} }\big] \big/ (3 \times 10^8),\\
    &{\tt Azimuth ~\ac{dod}} \coloneqq \Omega_{\mathrm{az}, n} \coloneqq \arctan\left( [y_{\mathrm{sc},n} - y_{\mathrm{rx}}]/[x_{\mathrm{sc},n} - x_{\mathrm{rx}} ] \right), \\
    &{\tt Azimuth ~\ac{doa}} \coloneqq  \Psi_{\mathrm{az},n} \coloneqq \arctan\left([ y_{\mathrm{rx}} - y_{\mathrm{sc},n} ]/ [x_{\mathrm{rx}} - x_{\mathrm{sc},n} ] \right), \\
    &{\tt Zenith ~\ac{dod}} \coloneqq  \Omega_{\mathrm{zn},n} \coloneqq \arctan\left([d_{\mathrm{2d}, {\bf r}_{\mathrm{sc},n} \rightarrow {\bf r}_{\mathrm{rx}}} ] / [z_{\mathrm{rx}} - z_{\mathrm{sc},n} ] \right) ,\\
    &{\tt Zenith ~\ac{doa}} \coloneqq \Psi_{\mathrm{zn},n} \coloneqq \arctan\left([d_{\mathrm{2d}, {\bf r}_{\mathrm{sc},n} \rightarrow {\bf r}_{\mathrm{rx}}} ]/ [ z_{\mathrm{sc},n} - z_{\mathrm{rx}} ] \right) ,
\end{align*}
where $d_{\mathrm{3d}, a \rightarrow b}$ and  $d_{\mathrm{2d}, a \rightarrow b}$ represents the $2$-D and $3$-D distance, respectively, between location $a$ and $b$.

\subsection{Problem Statement}


We consider the propagation channel a function of the RX's location.
As such, our focus here is on trajectory-based (or location-based) channel generation.
Concretely, given an initial trajectory and \ac{dd} channel evolution on that initial trajectory, we want to generate \ac{dd} channels for future trajectory points that must obey true channel statistics.
It is worth noting that this task is very similar to traditional \emph{time-series} analysis.
However, in our case, we do not consider this as a ``predictor" that must match the channel realizations for the future trajectory points. 
Instead, our goal is to generate \ac{dd} channel realizations in a way that the generated samples have the same statistics as the true realizations.
In order to generate the complete \ac{dd} channel from the \ac{dd} impulse response, we stack all \acp{mpc} information and prepare channel information vector $\mathbf{x}_t = \big\{{\bf r}_{\mathrm{rx}}, g, \{n, g_n, \tau_n, \Omega_{\mathrm{az},n}, \Omega_{\mathrm{zn},n}, \Psi_{\mathrm{az},n}, \Psi_{\mathrm{zn},n} \}_{n=1}^{N} \big\}$, where $g \coloneqq 10\log_{10}\left( \sum_{n=1}^N 10^{g_n/10} \right)$ and $\mathbf{x}_t \in \mathbb{R}^{4 + (N \times 7)}$.
Based on our above assumptions, $\mathbf{x}_t$ evolves along the  RX's trajectory\footnote{As we consider static scatterers and TX, there is no need to consider Doppler shift, see Sec. \ref{gscmSubSec}.}.



In this work, we essentially seek answer to the following question: \emph{given a sequence of \ac{dd} wireless channel information $\{\mathbf{x}_1, \mathbf{x}_2, \dots, \mathbf{x}_L \}$, where $L$ is the historical lag, can we generate complete \ac{dd} channel information for $\{\mathbf{x}_{L+1}, \mathbf{x}_{L+2}, \dots, \mathbf{x}_{L+P} \}$, where $P$ is the generation window, that follow certain channel statistics?}
These statistics are \ac{rms} delay and angular spreads.
It is worth noting that the term ``generative" modeling in this work refers to the fact that channel statistics are to be used to generate channel realizations for $P$ future trajectory points.

\section{Proposed Solution: Hybrid Transformer Model and Statistics-Aided Training Method}



We propose a \ac{htransformer} model and channel statistics-aided learning method in this section.
It is worth noting that other ML models can also be readily used for the proposed statistics-aided learning method.

\subsection{Proposed Hybrid-Transformer Architecture}

\begin{figure}[!t]
    \centering
    \includegraphics[trim=35 3 60 10,clip, width=0.49\textwidth,height=0.25\textheight]{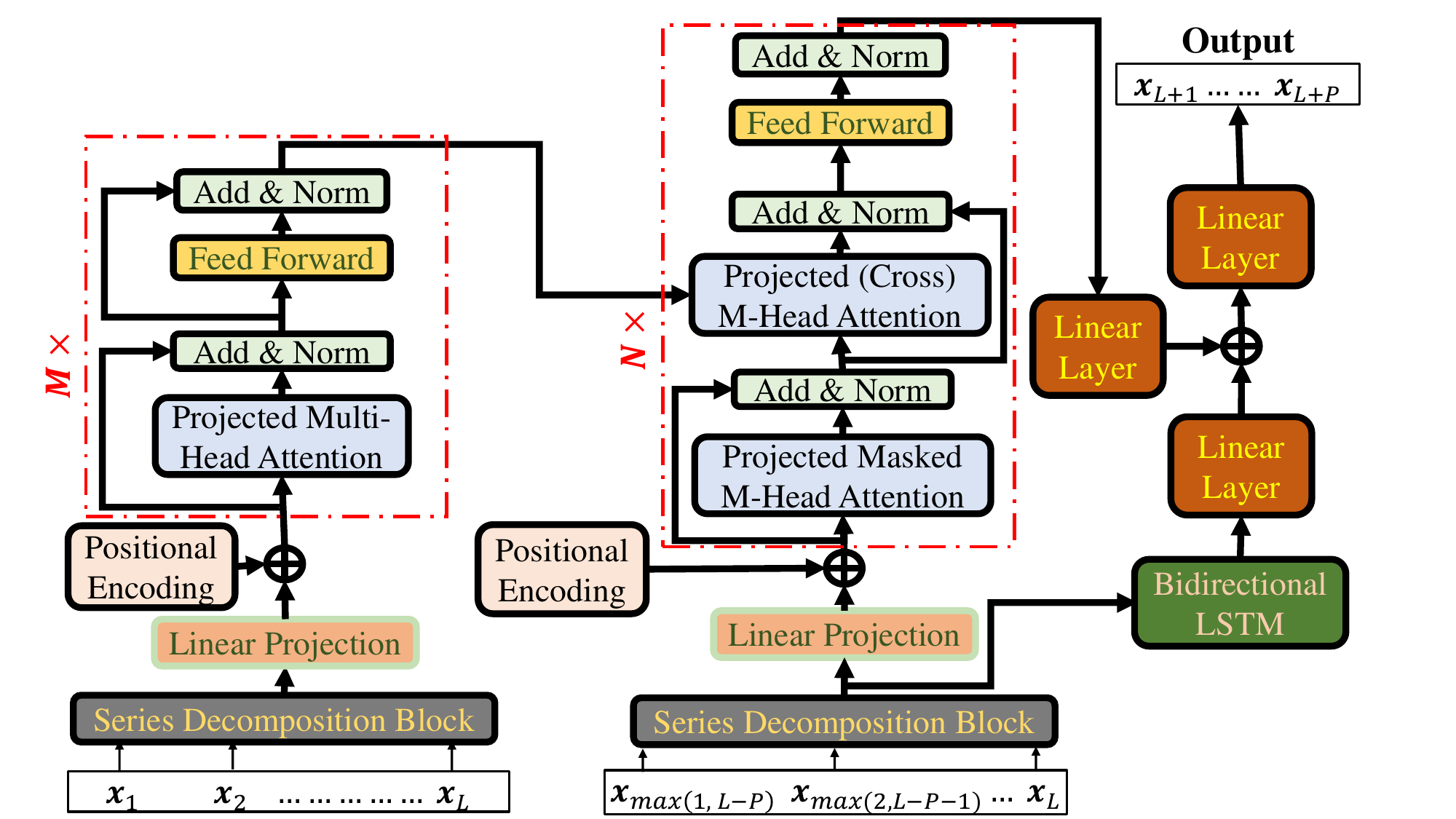}
    \caption{Proposed \ac{htransformer} model architecture}
    \label{hybridTransformerArchitecture}
\end{figure}

The proposed \ac{htransformer} model is built upon the original Transformer \cite{vaswani2017attention} and, thus, has an encoder-decoder architecture.
Fig. \ref{hybridTransformerArchitecture} shows the architecture of the proposed model.
We discuss the key components of the proposed model in what follows.


\subsubsection{Encoder}
\label{section_encoderside}
We first discuss the main building blocks of the encoder side.

\noindent
\textbf{Encoder Input}:
The encoder receives $L$ historical channel information, i.e., $\{\mathbf{x}_1, \mathbf{x}_2, \dots, \mathbf{x}_L \}$, as its input.

\noindent
\textbf{Series Decomposition Block (SDB)}:
The $L$ historical input first passes through the \ac{sdb} block, which returns the following as output.
 \begin{align}
    \mathbf{X}_{\mathrm{sdb,enc}} = \left\{ \mathrm{concat}\{\mathbf{x}_1, \mathbf{x}_1-\bar{\mathbf{x}}\}, \dots, \mathrm{concat} \{\mathbf{x}_L,\mathbf{x}_L - \bar{\mathbf{x}}\} \right\},
\end{align}
where $\bar{\mathbf{x}} \coloneqq \frac{1}{L}\sum_{l=1}^L{\mathbf{x}_l}$ and $\mathrm{concat}\{\mathbf{a} \in \mathbb{R}^F ,\mathbf{b} \in \mathbb{R}^F\} \coloneqq \mathrm{concat}\{[a_1, \dots, a_F], [b_1,\dots,b_F]\} = [a_1, \dots, a_F, b_1,\dots, b_F] \in \mathbb{R}^{2F}$.
Therefore, the \ac{sdb} increases the features of each channel information vector to $\mathbb{R}^{2 \times (4+(N+7))}$.
Since these additional features have zero mean, intuitively, these features may help the model to understand deviations from the mean.


\noindent
\textbf{Linear Projection and Positional Encoding}:  
The output of the \ac{sdb} then passes through a linear projected layer that converts each $\mathbb{R}^{2 \times (4+(N+7))}$ dimensional channel feature vector into a $\mathbb{R}^d_{\mathrm{model}}$ dimensional vector. 
It is worth noting that since our inputs are not words, we do not have any word \emph{embedding} as in the original Transformer, and a similar linear projection strategy is also widely used for image classification \cite{dosovitskiy2021an}.
To that end, the projected output gets added with positional encoding, which follows the procedure described in \cite{vaswani2017attention}. 
Let us denote the output of this addition by $\mathbf{X}_{\mathrm{pos\_enc}} \in \mathbb{R}^{L \times d_{\mathrm{model}}}$.  

\noindent
\textbf{Projected Self Multi-Head Attention}:
Given $\mathbf{X}_{\mathrm{pos\_enc}} \in \mathbb{R}^{L \times d_{\mathrm{model}}}$, we can calculate multi-head self-attention as \cite{vaswani2017attention} 
\begin{align}
    {\tt{MH}} \left( \mathbf{X}_{\mathrm{pos\_enc}} \right) = \mathrm{Concat} \left(\tt{head_1, head_2, \dots, head_h}\right) W_{\mathrm{O}},
\end{align}
where each ${\tt head_i}$ is calculated as
\begin{align}
   \rs\rs {\tt head_i} \rs&= \mathrm{Attention}\left(\mathbf{Q}_i, \mathbf{K}_i, \mathbf{V}_i \right) = \underbrace{\mathrm{softmax} \left[ \frac{\mathbf{Q}_i \left(\mathbf{K}_i\right)^T}{\sqrt{d_{\mathrm{model}}}} \right]}_{\mathbf{P} \in \mathbb{R}^{L \times L}} \mathbf{V}_i,
\end{align}
where $\mathbf{Q}_i = \mathbf{X}_{\mathrm{pos\_enc}} \mathbf{W}_{i,\mathrm{Q}}$, $\mathbf{K}_i = \mathbf{X}_{\mathrm{pos\_enc}} \mathbf{W}_{i,\mathrm{K}}$ and $\mathbf{V}_i = \mathbf{X}_{\mathrm{pos\_enc}} \mathbf{W}_{i,\mathrm{V}}$. Besides, $\mathbf{W}_{i,\mathrm{Q}} \in \mathbb{R}^{d_\mathrm{model} \times d_k}$, $\mathbf{W}_{i,\mathrm{K}} \in \mathbb{R}^{d_\mathrm{model} \times d_k}$, $\mathbf{W}_{i,\mathrm{V}} \in \mathbb{R}^{d_\mathrm{model} \times d_v}$ and $\mathbf{W}_{\mathrm{O}} \in \mathbb{R}^{h d_v \times d_\mathrm{model} }$ are learned matrices with hidden projection dimensions of $d_k$, $d_k=d_\mathrm{model}/h$ and $d_v$.
It is worth noting that the \emph{context mapping matrix $\mathbf{P}$} requires multiplying two $(L \times d_k)$ dimensional matrices, which has $\mathcal{O}(L^2)$ space and time complexity: the computation becomes prohibitive when $L$ is extremely large.

While there are different techniques to deal with this attention calculation, a recent study suggests that self-attention has low rank \cite{wang2020linformer}.
Motivated by this, we use a low-dimensional projection of the key and value matrices similar to \cite{wang2020linformer}. 
Concretely, we project $\mathbf{K}_i$ and $\mathbf{V}_i$ into $(B\times d_k)$ dimension, where $B \ll L$ as $\tilde{\mathbf{K}}_i = \mathbf{E}_i \mathbf{K}_i$, $\tilde{\mathbf{V}}_i = \mathbf{F}_i \mathbf{V}_i$, where $\mathbf{E}_i$ and $\mathbf{F}_i$ are $(B \times d_k)$ dimensional learned projection matrices.
Then, we calculate the projected attention as \cite{wang2020linformer} 
\begin{align}
    &\rs\rs{\tt \tilde{head}_i} \rs= \rs \mathrm{Attention}\left(\mathbf{Q}_i, \tilde{\mathbf{K}}_i, \tilde{\mathbf{V}}_i \right) = \underbrace{\mathrm{softmax} \left[ \frac{\mathbf{Q}_i \left(\tilde{\mathbf{K}}_i\right)^T}{\sqrt{d_{\mathrm{model}}}} \right]}_{\tilde{\mathbf{P}} \in \mathbb{R}^{L \times B}} \tilde{\mathbf{V}}_i,\\
    &{\tt \tilde{MH} } \left( \mathbf{X}_{\mathrm{pos\_enc}} \right) 
    = \mathrm{Concat} \left(\tt{\tilde{head}_1, \tilde{head}_2, \dots, \tilde{head}_h}\right) \mathbf{W}_{\mathrm{O}}.
\end{align}
This low-dimensional projection thus reduces the quadratic space and time complexity to a linear complexity of $\mathcal{O}\left(B \times L\right)$.

The rest of the components in the encoder side follow the original Transformer \cite{vaswani2017attention} architecture and, thus, are skipped in this paper for brevity.

\subsubsection{Decoder} 
Many of the encoder side building blocks are also used on the decoder side. 
The key differences are summarized below.

\noindent
\textbf{Decoder Inputs}:
While the traditional Transformer uses the same $L$ (shifted right) length sequence as input, our decoder side input depends on the length of the generation window $P$.
More specifically, the decoder side \ac{sdb} gets the last $\{\mathbf{x}_{L-P}, \dots, \mathbf{x}_{L}\}$ channel samples when $L>P$, while it sees all $L$ historical samples when $L \leq P$.
We represent the decoder side input succinctly by $\left\{\mathbf{x}_{\mathrm{max}\{1, L-P\}}, \mathbf{x}_{\mathrm{max}\{2, L-P-1\}}, \dots, \mathbf{x}_L \right\}$  .
It is worth noting that we are interested in the latter case, i.e., $L \ll P$.


\noindent
\textbf{Decoder Side \ac{sdb}}:
The decoder side \ac{sdb} operations depends on the generation window $P$.
Particularly, when $L \geq P$, the decoder side \ac{sdb} returns 
\begin{align}
    \mathbf{X}_{\mathrm{sdb,dec}} 
    =& \big\{ \mathrm{concat}\left\{\mathbf{x}_{\mathrm{max}\{1, L-P\} } , \mathbf{x}_{\mathrm{max}\{1, L-P\}} - \tilde{\mathbf{x}}\right\}, \dots, \nonumber\\ 
    &\qquad \mathrm{concat} \{\mathbf{x}_L,\mathbf{x}_L - \Tilde{\mathbf{x}}\} \big\},
\end{align}
where $\Tilde{\mathbf{x}} \coloneqq \frac{\mathbf{x}_{\mathrm{max}\{1, L-P\}} + \mathbf{x}_{\mathrm{max}\{2, L-P-1\}} + \dots + \mathbf{x}_L}{P}$.
However, when $P>L$, we used the available $L$ historical information and $(P-L)$ placeholders filled with the \emph{mean} and \emph{variance} information as  
\begin{align}
    \mathbf{X}_{\mathrm{sdb,dec}} 
    =&\bigg\{ \mathrm{concat}\{\mathbf{x}_1, \mathbf{x}_1-\bar{\mathbf{x}}\}, \dots, \mathrm{concat} \{\mathbf{x}_L,\mathbf{x}_L - \bar{\mathbf{x}} \}, \nonumber\\
    &\qquad \qquad \quad \underbrace{\mathrm{concat}\{ \bar{\mathbf{x}}, \breve{\mathbf{x}}\}, \dots, \mathrm{concat}\{ \bar{\mathbf{x}}, \breve{\mathbf{x}}\} }_{\mathrm{repeats} ~(P-L)~ \mathrm{times}} \bigg\},
\end{align}
where $\breve{\mathbf{x}} \coloneqq \frac{1}{L} \sum_{l=1}^L \left(\mathbf{x}_l - \bar{\mathbf{x}}\right)^2$ is the variance.


The projected masked multi-head attention is calculated using the same operations described above in Section \ref{section_encoderside} with an additional causal mask \cite{vaswani2017attention}.
Besides, the rest of the blocks inside the decoder also follow the respective block's architecture as in \cite{vaswani2017attention}.

\subsubsection{Hybrid Output Aggregations}

We leverage the advantages of the Transformer and \ac{bilstm} for the final generated output sequence.
Intuitively, we want to take the short-term dependency capturing benefits of \ac{lstm} and the Transformer's power to understand long sequences.
Besides, we used \ac{bilstm} instead of \ac{lstm} since \ac{bilstm} processes sequences in both forward and backward directions, which can provide some critical future contexts.
As such, we expect that integrating Transformer with \ac{bilstm} will capture both long-term and short-term dependencies.
More specifically, the decoder side \ac{sdb} output is fed to a \ac{bilstm} layer, and then the output of the decoder gets aggregated with the output of this \ac{bilstm} layer.
This aggregated output is then fed to a \ac{fc} layer that generates the final output sequence of \ac{dd} channel information $\hat{\mathbf{X}}_{\mathrm{out}} \coloneqq \{\mathbf{x}_{L+1}, \mathbf{x}_{L+2}, \dots, \mathbf{x}_{L+P} \} \in \mathbb{R}^{P \times (4 + (N \times 7))}$.

\subsection{Proposed Channel Statistics-Aided Training Method}

We want to utilize the \ac{dd} channel statistics to evaluate the generated channel realization $\hat{\mathbf{X}}_{\mathrm{out}}$. 
As such, we leverage domain knowledge to design a custom loss function $l \left(\mathbf{X}_{\mathrm{true}}, \hat{\mathbf{X}}_{\mathrm{out}}\right)$, which is defined as 
\begin{align}
\label{customLossEqn}
    &l \left(\mathbf{X}_{\mathrm{true}}, \hat{\mathbf{X}}_{\mathrm{out}}\right) \coloneqq l\left(S_\tau,\hat{S}_\tau\right) \alpha_\tau + \alpha_{\mathrm{az}} \big[l\left(S_{\Omega_{\mathrm{az}}},\hat{S}_{\Omega_{\mathrm{az}}}\right) + \nonumber\\
    &\qquad \qquad \qquad l\left(S_{\Psi_{\mathrm{az}}},\hat{S}_{\Psi_{\mathrm{az}}}\right) \big] + \alpha_{\mathrm{zn}} \big[ l\left(S_{\Omega_{\mathrm{zn}}},\hat{S}_{\Omega_{\mathrm{zn}}}\right) + \nonumber\\
    &\qquad \qquad \quad l\left(S_{\Psi_{\mathrm{zn}}},\hat{S}_{\Psi_{\mathrm{zn}}}\right) \big] + l \left(\{g_n\}_{n=1}^N, \{\hat{g}_n\}_{n=1}^N\right) \alpha_g,
\end{align}
where $\alpha_\tau$, $\alpha_{\mathrm{az}}$, $\alpha_{\mathrm{zn}}$ and $\alpha_g$ are weighing factor for the \ac{rms} delay spread, azimuth angular spreads, zenith angular spreads and \acp{mpc}' gains. 
It is worth noting that in order to calculate the statistics, we convert scaled $\mathbf{X}_{\mathrm{true}}$ and $\hat{\mathbf{X}}_{\mathrm{out}}$ back to original scales.
As such, these weighting factors are chosen in such a way that if those are multiplied by the corresponding true values, the result becomes $\approx 1$.
Moreover, $l(y, \hat{y})$ is ${\tt SmoothL1Loss}$, which often works better than mean squared error in the presence of outliers and is defined as
\begin{equation}
\label{smoothL1Loss}
\begin{aligned}
    l(y,\hat{y}) &=
    \begin{cases}
    0.5\left(y - \hat{y}\right)^2/\beta & \text{if } \left\vert y - \hat{y}\right\vert < \beta  \\
    \left\vert y-\hat{y} \right\vert - 0.5\beta, & \text{otherwise}
    \end{cases},
\end{aligned}
\end{equation}
where $\beta$ is the hyper-parameter that works as the threshold to transit between $L_1$ and $L_2$ losses.

Note that the calculations of the above statistics add additional computation overheads during the model training.
However, we can use matrix operations using appropriate indexing and calculate the statistics without any ``for" loop: the computation is relatively fast with modern \acp{gpu}. 
Besides, we assume the training happens offline; hence, these small overheads do not matter significantly for our application.
We stress that since our method emphasizes learning the \ac{dd} channel statistics, the generated realizations do not necessarily follow the ground truth channel realizations.
Therefore, while the difference in the statistics should match closely, the difference between the realizations can be significantly high.


\section{Simulation Settings and Discussions}

\subsection{Dataset Generation and Pre-Processing}
\noindent
\subsubsection{Dataset Generation}
We first generate a realistic trajectory of the RX assuming the TX is at ${\bf r}_{\mathrm{tx}}=(0,0,25)$, the RX's height is fixed $h_{\mathrm{rx}}=z_{\mathrm{rx}}=1.5$ and that the RX has initial starting point ${\bf r}_{\mathrm{rx}}^{t=0}$.
We then generate the RX's trajectory using 
\begin{align}
    x_{\mathrm{rx}} \gets x_{\mathrm{rx}} + \Delta_{\mathrm{2d}} \times \cos(\theta), \\
    y_{\mathrm{rx}} \gets y_{\mathrm{rx}} + \Delta_{\mathrm{2d}} \times \sin(\theta),
\end{align}
where $\theta \in \{\theta_1, \dots, \theta_A\}$ is a predefined set of heading angles.
These $A=50$ heading angles are chosen as
\begin{align}
   \rs \rs \rs \theta_i = 2\pi \times \sin \rs \left(0.1\pi + \frac{(i-1)}{A-1} (2\pi - 0.1\pi) \rs \right), ~ i = 1,\dots, A. 
\end{align} 
Besides, we assume that RX moves in the same heading angle $\theta$ for $\mathrm{H}$ consecutive steps, where $H$ is drawn from $[100,500]$ or if $d_{\mathrm{2d},{\bf r}_{\mathrm{tx} \rightarrow {\bf r}_{\mathrm{rx}}}} > 600$ meters.
Furthermore, we consider $N=26$ scatterers and their $x$, $y$ and $z$ coordinates are drawn uniformly randomly from $[-550, 500]$, $[-550, 500]$ and $[0, 30]$, respectively. 
Once the trajectory is known, we generate the \ac{gscm} channel information based on the equations described in Section \ref{gscmSubSec} with $f_c=2.4$ GHz.
In particular, we generate $125K$ total samples: $100K$ of these samples are used for model training and the rest $25K$ for model evaluation.

\subsubsection{Dataset Pre-Processing}

\noindent
In our raw data, the delays, angle, and power information are in nanoseconds, degrees, and dBm. 
We then use a customized \emph{min-max} scaler that scales each data feature $x_{t,f} \in \mathbf{x}_t$ that are not fixed, as  
\begin{equation}
    x_{t,f,\mathrm{scaled}} = \big[x_{t,f} - x_{t,f,\mathrm{min}}\big] \big/ \big[x_{t,f,\mathrm{max}} - x_{t,f,\mathrm{min}}\big], 
\end{equation}
where $x_{t,f,\mathrm{max}}$ and $x_{t,f,\mathrm{min}}$ are the maximum and minimum values of the the $f^{\mathrm{th}}$ feature from the entire training dataset. 
Besides, the fixed features: $\{h_{\mathrm{rx}}, \{n\}_{n=1}^N\}$, are scaled by dividing by $N$.
Note that these features are set to zeros in traditional min-max scalar. 
However, since we stack all \acp{mpc}' features sequentially to prepare each channel information vector $\mathbf{x}_t$, setting each (constant) path ID to $0$ may not help the model distinguish features from different paths.

\subsection{Key Simulation Parameters}

The \ac{htransformer} model has $h=8$ heads and 2 encoder / decoder / \ac{bilstm} layers.
The \ac{bilstm} hidden size is $128$ and $d_{\mathrm{model}}=512$.
Besides, $512$ \ac{fc} layers in the position-wise feed forward blocks of the encoder and decoder, low rank projection dimension $B=64$, batch size of $256$, PyTorch\footnote{\url{https://pytorch.org/}} module with {\tt AdamW} optimizer with initial learning rate of $5\times 10^{-5}$, which is linearly decayed by $10\%$ in every $10$ training round, and $\beta=1$ is used.
We trained the model for $250$ episodes on HPC cluster using NVIDIA A$40$ and A$100$ \acp{gpu}.

\begin{figure*}[!t]
\begin{subfigure}{0.33\textwidth}
    \centering
    \includegraphics[trim=15 12 10 15, clip, width=\textwidth, height=0.16\textheight]{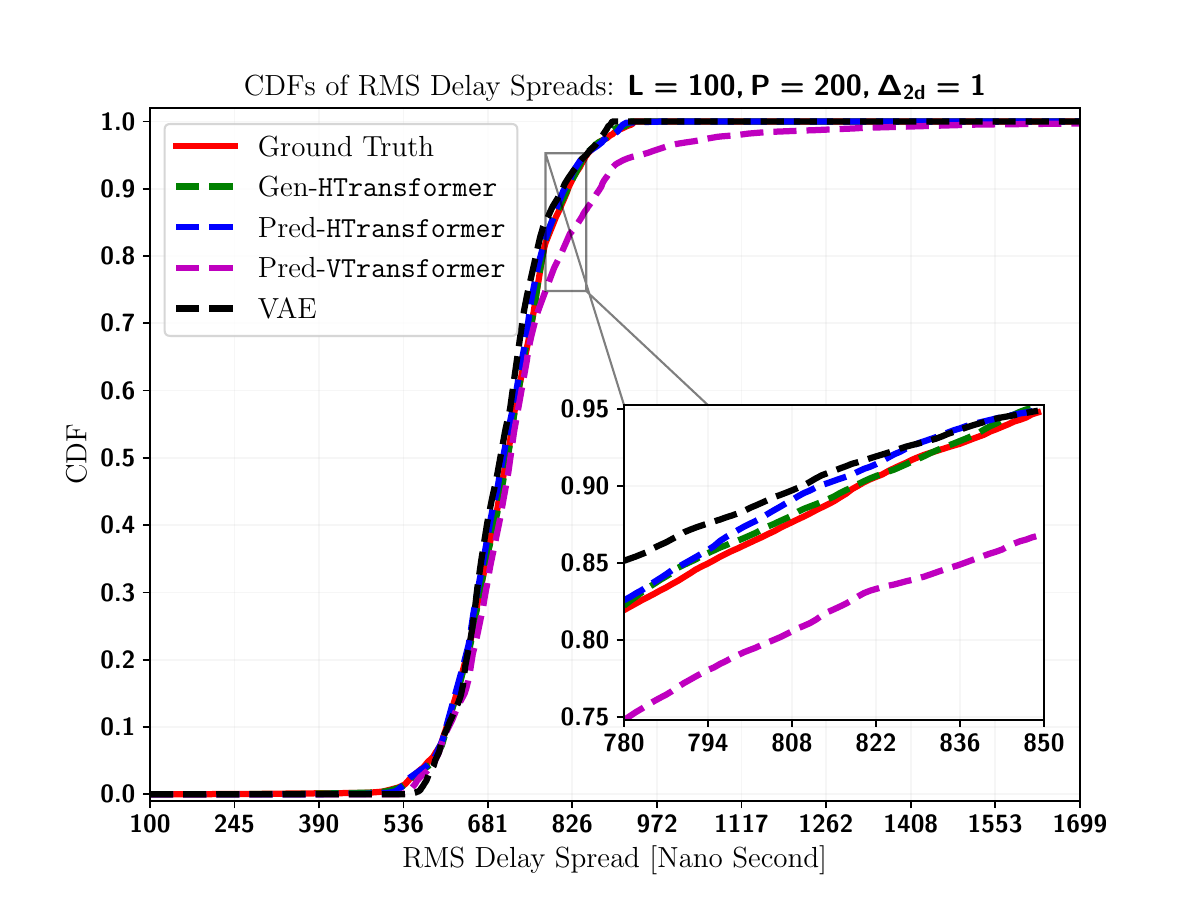}
    \caption{CDF of \ac{rms} delay spread: $P=200$ }
    \label{rmsDelay_100}
\end{subfigure}
\begin{subfigure}{0.33\textwidth}
    \centering
    \includegraphics[trim=15 12 10 15, clip, width=\textwidth, height=0.16\textheight]{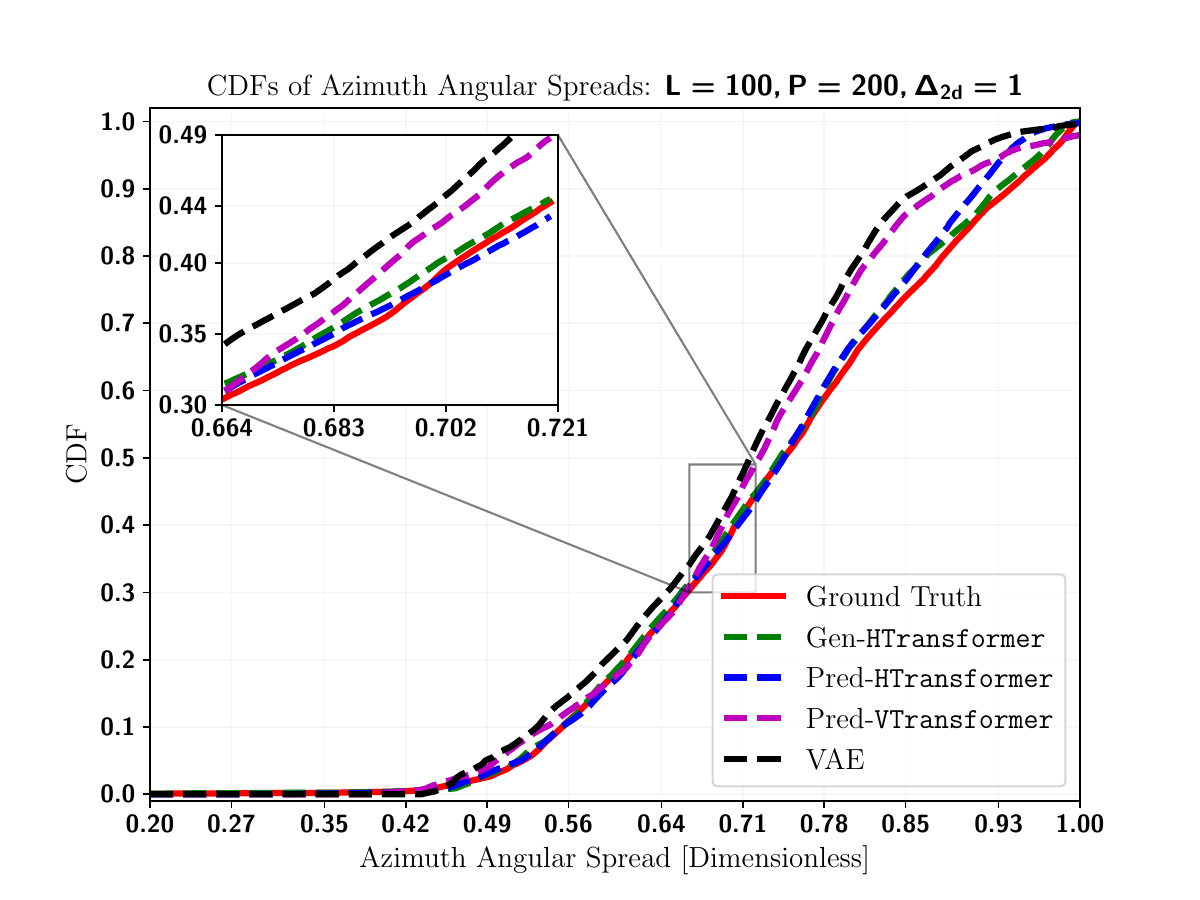}
    \caption{CDF of azimuth \ac{dod} spread: $P=200$}
    \label{angularspreadAz_100}
\end{subfigure}
\begin{subfigure}{0.33\textwidth}
\centering
    \includegraphics[trim=15 12 10 15, clip, width=\textwidth, height=0.16\textheight]{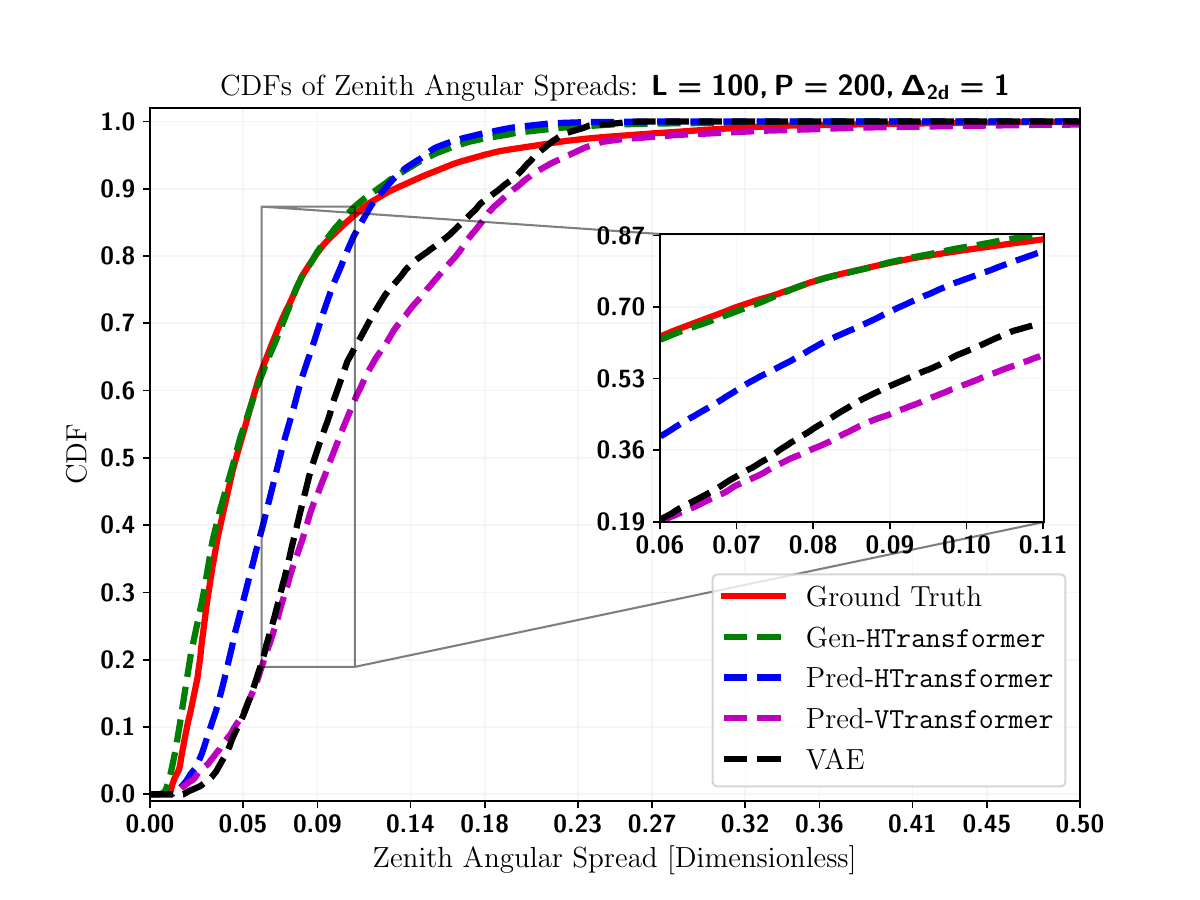}
    \caption{CDF of zenith \ac{dod} spread: $P=200$}  
    \label{angularspreadZn_100}
\end{subfigure}
\begin{subfigure}{0.33\textwidth}
    \centering
    \includegraphics[trim=15 12 10 15, clip, width=\textwidth, height=0.16\textheight]{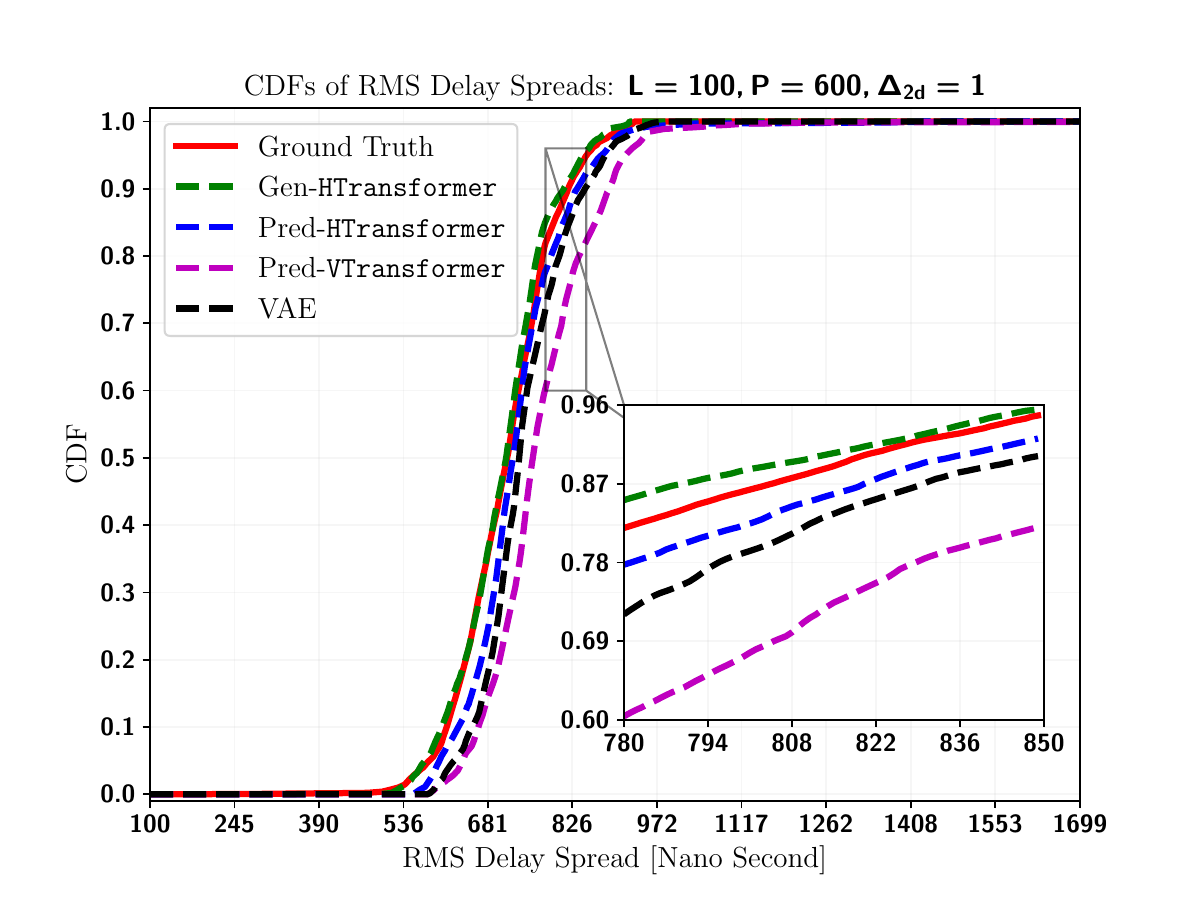}
    \caption{CDF of \ac{rms} delay spread: $P=600$ }
    \label{rmsDelay_600}
\end{subfigure}
\begin{subfigure}{0.33\textwidth}
    \centering
    \includegraphics[trim=15 12 10 15, clip, width=\textwidth, height=0.16\textheight]{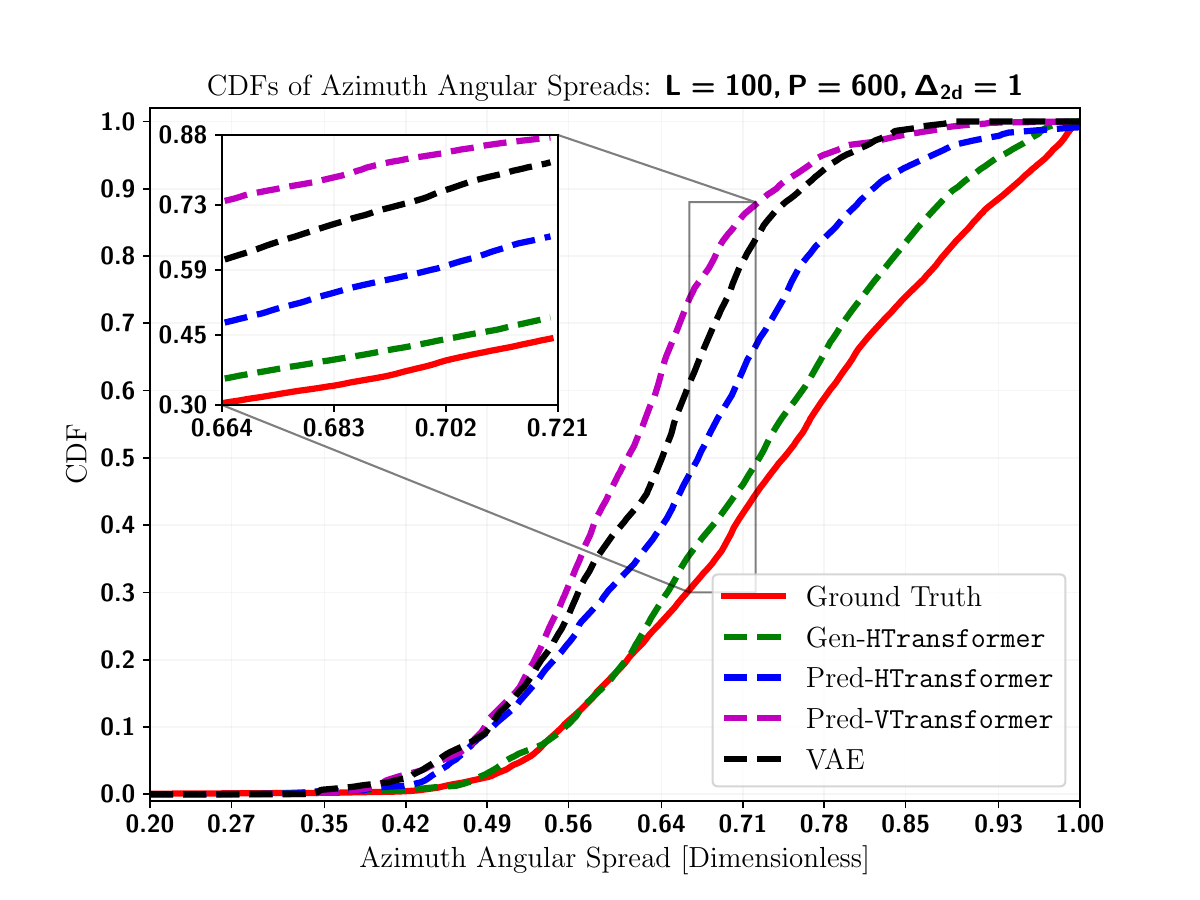}
    \caption{CDF of azimuth \ac{dod} spread: $P=600$}
    \label{angularspreadAz_600}
\end{subfigure}
\begin{subfigure}{0.33\textwidth}
\centering
    \includegraphics[trim=15 12 10 15, clip, width=\textwidth, height=0.16\textheight]{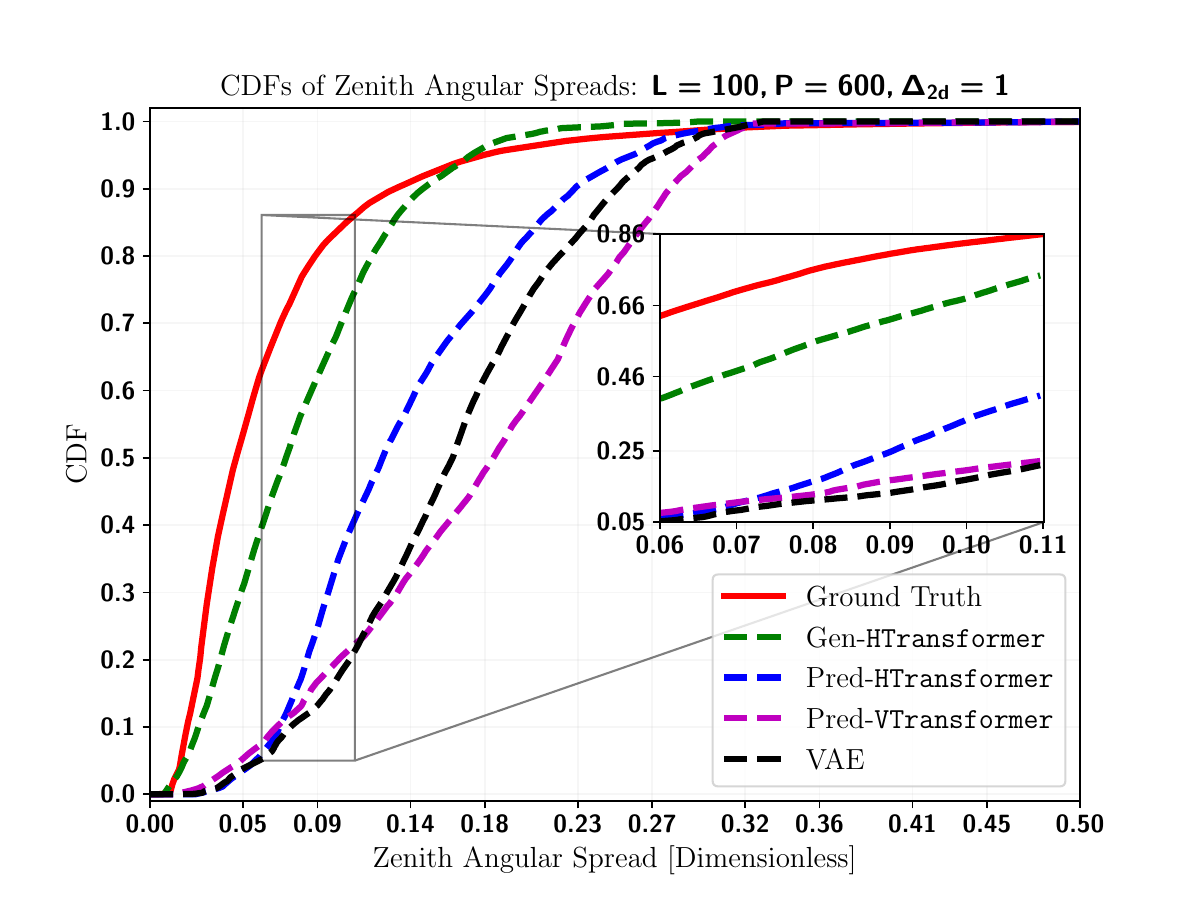}
    \caption{CDF of zenith \ac{dod} spread: $P=600$}  
    \label{angularspreadZn_600}
\end{subfigure}
\caption{\bblue{Test performance: delay and angular spreads' distributions when $L=100$ (from $3$ independent runs)}}
\label{delayAngleFigs}
\end{figure*}
\begin{figure}[!t]
\begin{subfigure}{0.23\textwidth}
\centering
    \includegraphics[trim=15 12 10 15, clip, width=\textwidth]{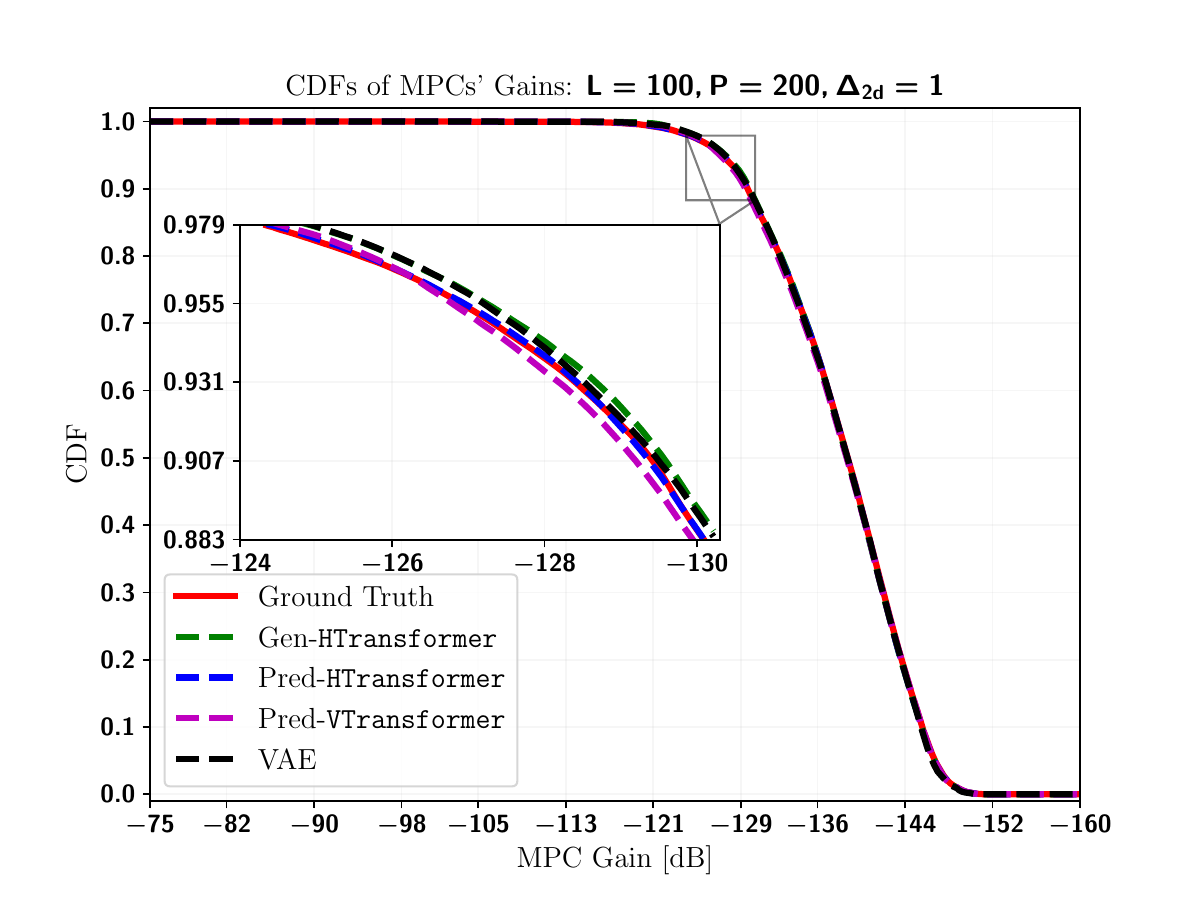}
    \caption{CDF of \ac{mpc}'s power distribution: $P=200$}  
    \label{mpcPowerDistribution_100}
\end{subfigure}
\begin{subfigure}{0.23\textwidth}
\centering
    \includegraphics[trim=15 12 10 15, clip, width=\textwidth]{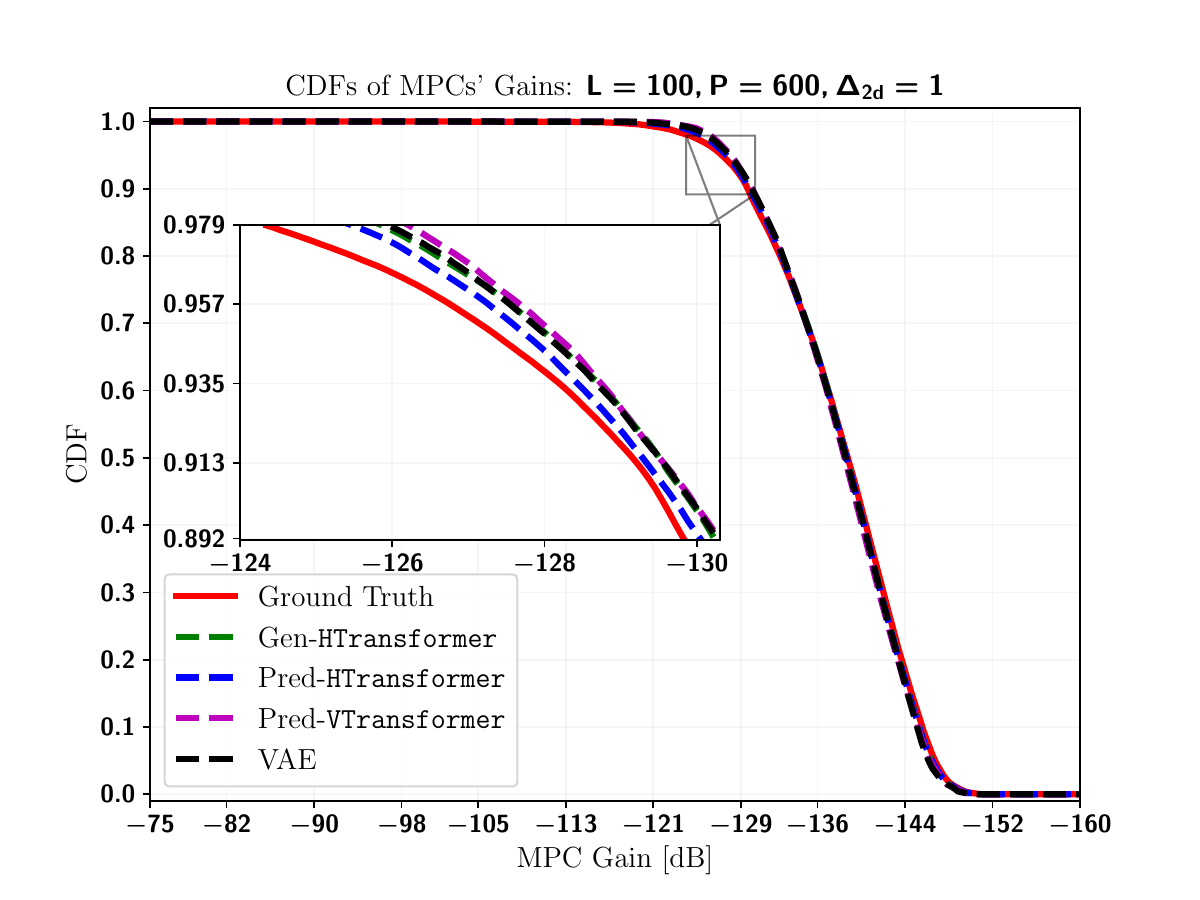}
    \caption{CDF of \ac{mpc}'s power distribution: $P=600$}  
    \label{mpcPowerDistribution_600}
\end{subfigure}
\caption{\bblue{CDF of \ac{mpc}'s power distribution: $L=100$}}
\label{mpcPowerDist}
\end{figure}

\subsection{Performance Evaluation}

For performance evaluation, we consider \bblue{three} baselines: (a) our \ac{htransformer} with predictive tasks, termed as Pred-\ac{htransformer}, b) vanilla Transformer models that were used in \cite{zhou2024transformer,jiang2022accurate}, termed as Pred-{\tt VTransformer}\bblue{, and (c) a \ac{vae} model \cite{kingma2014auto}.  
While the {\tt Transformer} baselines calculate the ${\tt SmoothL1Loss}$ loss $l \left(\mathbf{X}_{\mathrm{true}}, \hat{\mathbf{X}}_{\mathrm{out}}\right)$ directly using (\ref{smoothL1Loss}), the \ac{vae}'s loss function follows \cite{kingma2014auto}.  
The encoder of the \ac{vae} has $512$ and $32$ hidden layers and latent space dimensions, respectively. 
Besides, for the \ac{vae}, we consider the $\left\{\mathbf{x}_1, \dots, \mathbf{x}_L, \mathbf{x}_{L+1}, \dots, \mathbf{x}_{L+P} \right\}$ as our original channel information vector that we want to reconstruct with observations $\left\{\mathbf{x}_1, \dots, \mathbf{x}_L, \mathbf{0}, \dots, \mathbf{0} \right\}$ that has only zeros in the last $P$ positions instead of $\mathbf{x}_{L+1}, \dots \mathbf{x}_{L+P}$. 
Moreover,} our proposed statistics-aided learning, termed as Gen-\ac{htransformer} uses (\ref{customLossEqn}) to evaluate training performance.


We first investigate whether Transformer-based models accurately capture channel statistics for different generation window $P$ with $\Delta_{2\mathrm{d}} = 1$ meter.
Since \ac{htransformer} is trained to minimize the loss between the statistics, we expect that Gen-{\tt Transformer} produces \ac{dd} channel samples that closely obey the ground truth statistics regardless of the generation window $P$.
Moreover, in the predictive case, since our proposed \ac{htransformer} model utilizes \ac{sdb} and \ac{bilstm}, we also expect it to understand data trends quite closely, i.e., the predicted channel realizations should closely match the true realizations. 
As such, we expect the statistics from Gen-\ac{htransformer} and Pred-\ac{htransformer} to align well with the ground truth.
On the other hand, Pred-{\tt vTransformer} may fail to capture the small-scale trends but still learn some long-term trends --- thanks to its attention mechanism --- that help it to predict the channel realizations with different statistics.

We indeed observe these trends in our simulation results, as shown in Figs. \ref{delayAngleFigs} - \ref{mpcPowerDist}.
\bblue{Note that all results are the average of three independent trials.} 
In particular, we observe that the RMS delay and angular spreads match quite closely with our proposed \ac{htransformer} model.
Besides, as $P$ increases from $200$ to $600$, Gen-\ac{htransformer} gets better at matching the statistics. 
On the other hand, the statistics from the existing Pred-{\tt vTransformer} deviate from the ground truth.
\bblue{For example, the \ac{mse} between the ground truth \ac{cdf} and Gen-\ac{htransformer} generated realizations' \ac{cdf} for the \ac{rms} delay spread at $P=600$ is only $-40.2$ dB.
The corresponding differences for Pred-\ac{htransformer}, Pred-{\tt vTransformer}, and \ac{vae} are $-32.12$ dB, $-22.07$ dB, and $-26.87$ dB, respectively. 
Besides, the corresponding \ac{mse} differences for the \ac{rms} azimuth angular spreads are $-27.54$ dB, $-17.68$ dB, $-12.80$ dB, and $-14.17$ dB.
The benefit of Gen-\ac{htransformer} is also evident for the \ac{rms} zenith angular spread, with \ac{mse} difference of only $-21.61$ dB for $P=600$, whereas the three baselines respectively have $-12.16$ dB, $-9.44$ dB, and $-10.07$ dB differences.}  
Moreover, the distribution of \acp{mpc} powers in Fig. \ref{mpcPowerDist} is close to the ground truth in all cases.
This is due to the fact that the gains are calculated based on the path loss equation that entirely depends on the distance (the other parameters are fixed in our case).


\begin{figure}
\centering
    \includegraphics[trim=15 12 10 15, clip, width=0.95\linewidth]{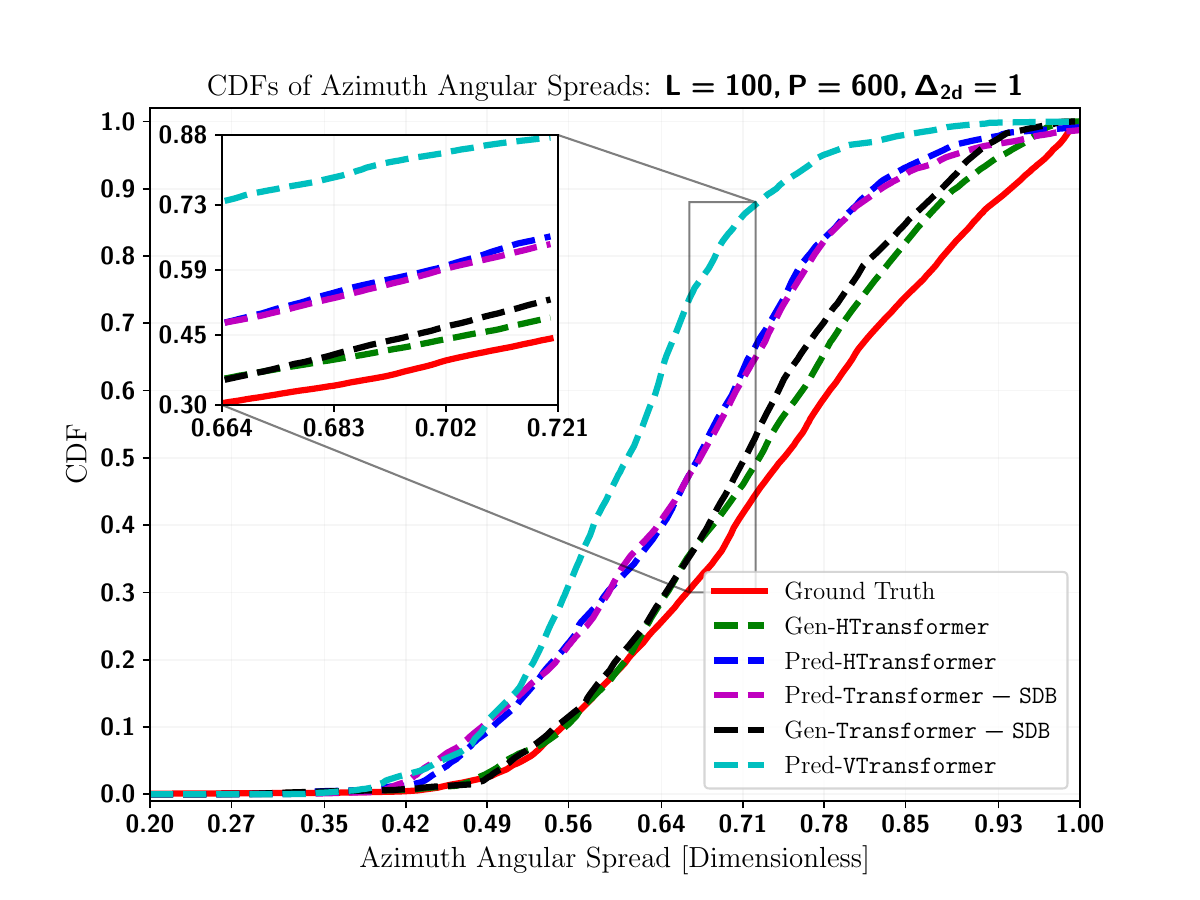}
    \caption{\bblue{CDF of MPC's azimuth angular spreads with different {\tt Transformer} architectures}}
    \label{cdfComp_Trans_SDB_LSTM_Comp}
\end{figure}

\bblue{Now, we show the impact of different component blocks of the proposed \ac{htransformer} model. 
To do that, we drop the hybrid output aggregation, i.e., \ac{bilstm} block, while keeping the \ac{sdb} intact. We call this {\tt Transformer-\ac{sdb}} and train it following (a) the predictive case (regular loss function) and (b) the generative case (custom loss function as in (\ref{customLossEqn})).  
Besides, when we also drop the \ac{sdb} block, the architecture boils down to the architecture of the Pred-{\tt vTransformer}. 
From Fig. \ref{cdfComp_Trans_SDB_LSTM_Comp}, it is clear that without the \ac{bilstm} model block, neither the predictive nor the statistics-aided generative solution achieves the performance of the proposed Gen-\ac{htransformer}. 
For example, the {\tt Transformer-SDB} has the \ac{mse} difference of $-17.95$ dB and $-24.22$ dB with regular and statistics-aided loss functions, respectively, while our proposed Gen-\ac{htransformer} has a difference of $-27.54$ dB. This behavior is consistent with other \ac{mpc} parameters, particularly when $P$ is very large.}

To that end, we want to investigate how well the proposed generative solution works for different $P$ and $\Delta_{2\mathrm{d}}$.
Intuitively, our method should work well regardless of $P$ and $\Delta_{2\mathrm{d}}$ since the training method inherently considers the differences of statistics as the performance evaluation criteria.
Indeed, the listed simulation results in Tables \ref{rmsDelaySpreadDistributionComp}-\ref{gainDistributionComp} validate this. 
We see a small mean square error in the generated and true distributions for different generation lengths and $\Delta_{2\mathrm{d}}$.

\begin{table}[!t]
\centering
\caption{Mean squared difference of RMS delay spread distributions (from $3$ independent runs): lag, $L=100$}
\begin{tabular}{|>{\centering \arraybackslash}m{1.2 cm}|>{\centering\arraybackslash}m{1.8cm}|>{\centering\arraybackslash}m{1.8cm}|>{\centering\arraybackslash}m{1.8cm}|}
    \hline
    \rowcolor[HTML]{CCFFCC} 
    \textbf{Generation Length} $P$ & \textbf{Difference:} $\Delta_{2\mathrm{d}}=0.5$ m & \textbf{Difference} $\Delta_{2\mathrm{d}}=1$ m & \textbf{Difference:} $\Delta_{2\mathrm{d}}=1.5$ m \\ \hline
    \rowcolor[HTML]{F2F2F2} 
    $P=100$& $-52.8885$ dB & $-55.2699$ dB & $-54.5774$ dB \\ \hline
    \rowcolor[HTML]{E6E6E6} 
    $P=200$ & $-49.5570$ dB & $-53.9503$ dB & $-52.1240$ dB\\ \hline
    \rowcolor[HTML]{F9F9F9} 
     $P=300$ & $-47.0285$ dB & $-48.6410$ dB &$-49.2194$ dB \\ \hline
    \rowcolor[HTML]{E8E8E8} 
     $P=400$ & $-41.5335$ dB & $-44.3583$ dB& $-44.3706$ dB \\ \hline
    \rowcolor[HTML]{FAFAFA} 
    $P=500$ & $-42.3484$ dB & $-33.5980$ dB & $-41.4754$ dB \\ \hline
    \rowcolor[HTML]{F0F0F0} 
    $P=600$ & $-38.7279$ dB & $-40.1955$ dB & $-39.7902$ dB\\ \hline
\end{tabular}
\label{rmsDelaySpreadDistributionComp}
\end{table}
\begin{table}[!t]
\caption{Mean squared difference of azimuth angular spread distributions (from $3$ independent runs): lag, $L=100$}
\centering
\begin{tabular}{|>{\centering \arraybackslash}m{1.2 cm}|>{\centering\arraybackslash}m{1.8cm}|>{\centering\arraybackslash}m{1.8cm}|>{\centering\arraybackslash}m{1.8cm}|}
    \hline
    \rowcolor[HTML]{CCFFCC} 
    \textbf{Generation Length} $P$ & \textbf{Difference:} $\Delta_{2\mathrm{d}}=0.5$ m & \textbf{Difference} $\Delta_{2\mathrm{d}}=1$ m & \textbf{Difference:} $\Delta_{2\mathrm{d}}=1.5$ m \\ \hline
    \rowcolor[HTML]{F2F2F2} 
    $P=100$& $-42.0474$ dB & $-38.7102$ dB & $-38.5716$ dB\\ \hline
    \rowcolor[HTML]{E6E6E6} 
    $P=200$ & $-42.7066$ dB & $-39.2833$ dB & $-37.2502$ dB \\ \hline
    \rowcolor[HTML]{F9F9F9} 
     $P=300$ & $-40.1472$ dB & $-36.3108$ dB & $-31.2296$ dB \\ \hline
    \rowcolor[HTML]{E8E8E8} 
     $P=400$ & $-37.8697$ dB & $-30.0874$ dB &$-28.2045$ dB  \\ \hline
    \rowcolor[HTML]{FAFAFA} 
    $P=500$ & $-33.4387$ dB & $-27.7246$ dB & $-26.1974$ dB \\ \hline
    \rowcolor[HTML]{F0F0F0} 
    $P=600$ & $-30.4956$ dB & $-27.5395$ dB & $-24.9633$ dB \\ \hline
\end{tabular}
\label{azAngularSpreadDistributionComp}
\end{table}
\begin{table}[!t]
\centering
\caption{Mean squared difference of zenith angular spread distributions (from $3$ independent runs): lag, $L=100$}
\begin{tabular}{|>{\centering \arraybackslash}m{1.2 cm}|>{\centering\arraybackslash}m{1.8cm}|>{\centering\arraybackslash}m{1.8cm}|>{\centering\arraybackslash}m{1.8cm}|}
    \hline
    \rowcolor[HTML]{CCFFCC} 
    \textbf{Generation Length} $P$ & \textbf{Difference:} $\Delta_{2\mathrm{d}}=0.5$ m & \textbf{Difference} $\Delta_{2\mathrm{d}}=1$ m & \textbf{Difference:} $\Delta_{2\mathrm{d}}=1.5$ m \\ \hline
    \rowcolor[HTML]{F2F2F2} 
    $P=100$& $-35.6576$ dB & $-34.4315$ dB & $-32.2137$ dB \\ \hline
    \rowcolor[HTML]{E6E6E6} 
    $P=200$ & $-33.8098$ dB & $-34.5514$ dB & $-34.0743$ dB\\ \hline
    \rowcolor[HTML]{F9F9F9} 
     $P=300$ & $-33.1182$ dB & $-30.6249$ dB & $-28.1075$ dB \\ \hline
    \rowcolor[HTML]{E8E8E8} 
     $P=400$ & $-32.1355$ dB & $-28.8586$ dB & $-25.9413$ dB \\ \hline
    \rowcolor[HTML]{FAFAFA} 
    $P=500$ & $-33.0336$ dB & $-18.5252$ dB & $-23.8947$ dB \\ \hline
    \rowcolor[HTML]{F0F0F0} 
    $P=600$ & $-27.7930$ dB & $-21.6128$ dB & $-19.9344$ dB \\ \hline
\end{tabular}
\label{zenAngularSpreadDistributionComp}
\end{table}
\begin{table}[!t]
\caption{Mean squared difference of \ac{mpc} power distributions (from $3$ independent runs): lag, $L=100$}
\centering
\begin{tabular}{|>{\centering \arraybackslash}m{1.2 cm}|>{\centering\arraybackslash}m{1.8cm}|>{\centering\arraybackslash}m{1.8cm}|>{\centering\arraybackslash}m{1.8cm}|}
    \hline
    \rowcolor[HTML]{CCFFCC} 
    \textbf{Generation Length} $P$ & \textbf{Difference:} $\Delta_{2\mathrm{d}}=0.5$ m & \textbf{Difference} $\Delta_{2\mathrm{d}}=1$ m & \textbf{Difference:} $\Delta_{2\mathrm{d}}=1.5$ m \\ \hline
    \rowcolor[HTML]{F2F2F2} 
    $P=100$ & $-56.3837$ dB & $-57.5637$ dB & $-58.1858$ dB \\ \hline
    \rowcolor[HTML]{E6E6E6} 
    $P=200$ & $-53.9518$ dB & $-53.3001$ dB & $-54.2991$ dB\\ \hline
    \rowcolor[HTML]{F9F9F9} 
     $P=300$ & $-53.2143$ dB & $-53.1760$ dB & $-51.0421$ dB \\ \hline
    \rowcolor[HTML]{E8E8E8} 
     $P=400$ & $-51.4039$ dB & $-48.2488$ dB & $-47.1207$ dB \\ \hline
    \rowcolor[HTML]{FAFAFA} 
    $P=500$ & $-48.9298$ dB & $-40.6194$ dB & $-42.5530$ dB \\ \hline
    \rowcolor[HTML]{F0F0F0} 
    $P=600$ & $-46.3882$ dB & $-44.8855$ dB & $-41.4754$ dB\\ \hline
\end{tabular}
\label{gainDistributionComp}
\end{table}

\section{Conclusions}

We proposed a new \ac{htransformer} model and channel statistics-based training method to generate complete \ac{dd} channel realizations that match the true statistics.
Our extensive results suggest that if the channel is modeled as a function of (RX) location, the statistics-aided learning method yields a small performance difference even when the generation window is very long.


\bibliographystyle{IEEEtran}
\bibliography{referece.bib}

\end{document}